\documentstyle[12pt,epsfig]{article}

\newcommand{\be}{\begin{equation}}
\newcommand{\ee}{\end{equation}}

\newcommand{\dlt}{\delta}

\newcommand{\bS}{{\bf S}}

\newcommand{\bB}{{\bf B}}

\newcommand{\bt}{\beta}

\newcommand{\al}{\alpha}

\newcommand{\gm}{\gamma}
\newcommand{\om}{\omega}

\newcommand{\Gm}{\Gamma}

\begin{document}

\begin{center}

{\Large{\bf Collective spin dynamics in magnetic nanomaterials} \\ [5mm]

V.K. Henner$^{1,2}$, V.I. Yukalov$^3$, P.V. Kharebov$^1$,
and E.P. Yukalova$^4$} \\ [3mm]

{\it $^1$Perm State University, Perm 614000, Russia \\ [2mm]
$^2$University of Louisville, Louisville, Kentucky 40292, USA \\ [2mm]
$^3$Bogolubov Laboratory of Theoretical Physics, \\
Joint Institute for Nuclear Research, Dubna 141980, Russia \\ [2mm]
$^4$Laboratory of Information Technologies, \\
Joint Institute for Nuclear Research, Dubna 141980, Russia}

\end{center}

\vskip 1cm

{\bf E-mail:} yukalov@theor.jinr.ru

\begin{abstract}
Magnetic nanomaterials are considered, formed by magnetic nanomolecules
with high spins. The problem of spin reversal in these materials is
analyzed, which is of interest for the possible use of such materials
for quantum information processing and quantum computing. The fastest
spin reversal can be achieved by coupling the spin sample to a resonant
electric circuit and by an appropriate choice of the system parameters.
A principal point is to choose these parameters so that to organize
coherent spin motion. Dynamics of collective motion is modelled by
computer simulations, which confirm the high level of dynamical
coherence of molecular spins in the process of spin reversal.

\end{abstract}

\vskip 10mm

\section{Introduction}

There exists a wide class of nanomolecules possessing large magnetic
moments, reaching tens of the Bohr magnetons. Ensembles of such molecules
form crystalline clusters, called molecular nanomagnets. The properties
of these materials have been described in review articles [1--4]. Because
of the high molecular spins and regular crystalline structure, molecular
manomagnets are considered as promising candidates for quantum information
processing and quantum computing. The main problem, hindering their
application for that purpose, is a very slow relaxation of the magnetic
moment. While, for serving as a fast device for information processing,
it is necessary that magnetic moments could be quickly manipulated between
up and down directions.

In order to realize the spin reversal in molecular nanomagnets as an
ultrafast process, a mechanism has been suggested [3--7] allowing one to
reach the characteristic reversal time as short as $10^{-12}$ s. This can
be achieved by connecting the molecular sample to a resonant electric
circuit, providing feedback to the spin motion and organizing the spin
relaxation as a collective process with a high level of coherence in spin
dynamics.

There also exists another type of magnetic nanomaterials, formed by
magnetic nanoclusters [3], whose spins can reach $s\sim 100-10^4$. In
principle, ensembles of such nanoclusters could also be employed for
realizing collective spin relaxation as an ultrafast process. There is,
however, an important difference between nanomolecules and nanoclusters.
A collection of nanomolecules forms a well-defined crystalline structure,
with all molecules being exactly the same, possessing identical sizes and
all properties. But nanoclusters do not form crystals, their collections
representing a kind of powder. The most important is that it is impossible
to produce a large number of nanoclusters with identical properties. Sizes,
shapes, spins, and all other parameters of magnetic nanoclusters vary in a
wide range. A system of such different nanoclusters would form a strongly
nonuniform matter, with a large inhomogeneous broadening, hindering the
possibility of coherent spin motion. This is why magnetic nanomolecules
are preferable for realizing coherent spin dynamics.

The aim of the present communication is to study spin dynamics of magnetic
nanomolecules in the regime of their maximally fast spin reversal, which can
be found by varying the system parameters. We concentrate our attention on
the analysis of dynamical coherence of moving spins. For this purpose, we
study the temporal behavior of the coherence factor. We demonstrate that,
appropriately choosing the system parameters and tuning the Zeeman frequency
of nanomolecules in resonance with the electric circuit, it is feasible to
realize highly coherent spin dynamics.

\section{Basic equations}

The system of magnetic molecules is characterized by the Hamiltonian
\be
\label{1}
\hat H = \sum_i \hat H_i \; + \;
\frac{1}{2} \; \sum_{i\neq j} \hat H_{ij} \; ,
\ee
in which the index $i=1,2,\ldots, N$ enumerates molecules,
\be
\label{2}
\hat H_i = -\mu_0 \bB \cdot\bS_i - D (S_i^z)^2 \; ,
\ee
with $\mu_0=-2\mu_B$ being the electronic magnetic moment and $D$ being the
magnetic anisotropy parameter, the total magnetic field
\be
\label{3}
\bB = B_0 {\bf e}_z + H {\bf e}_x
\ee
includes an external static magnetic field $B_0$ and the resonator feedback
field $H$; the interaction term is
\be
\label{4}
\hat H_{ij} = \sum_{\al\bt} D_{ij}^{\al\bt} S_i^\al S_j^\bt \; ,
\ee
with the dipolar tensor
\be
\label{5}
D_{ij}^{\al\bt} = \frac{\mu_0^2}{r_{ij}^3} \;
\left ( \dlt_{\al\bt} - 3 n_{ij}^\al n_{ij}^\bt \right ) \; ,
\ee
where $r_{ij}\equiv|{\bf r}_{ij}|$, ${\bf r}_{ij}={\bf r}_{i}-{\bf r}_{j}$,
${\bf n}_{ij}\equiv{\bf r}_{ij}/r_{ij}$. The resonator feedback field is produced
by a magnetic coil surrounding the sample and can be described [8,9] by the
Kirchhoff equation
\be
\label{6}
\frac{dH}{dt} + 2 \gm H + \om^2 \int_0^t \; H(t')\; dt' =  -
4\pi\eta\; \frac{dm_x}{dt} \; ,
\ee
in which $\gm$ is the resonator damping, $\om$ is the resonator natural
frequency, $\eta$ is a filling factor, and
\be
\label{7}
m_x = \frac{\mu_0}{V} \; \sum_j < S_j^x>
\ee
is the transverse magnetization density, with $V$ being the sample volume.
The overall experimental setup is explained in detail in the review article
[4].

The main parameters characterizing the system are the Zeeman frequency
\be
\label{8}
\om_0 \equiv -\; \frac{\mu_0}{\hbar} \; B_0 =
\frac{2}{\hbar}\; \mu_B B_0 \; ,
\ee
the resonator natural frequency $\om$, the anisotropy frequency
\be
\label{9}
\om_D \equiv ( 2S-1) \; \frac{D}{\hbar} \; ,
\ee
the spin-phonon attenuation $\gm_1\equiv1/T_1$, the spin-dephasing
attenuation $\gm_2\equiv 1/T_2$, and the resonator dumping $\gm$. The
frequencies are assumed to satisfy the resonance conditions
\be
\label{10}
\left | \frac{\om-\om_0}{\om_0} \right | \; \ll \; 1 \; , \qquad
\frac{\om_D}{\om_0} \; \ll \; 1 \; .
\ee

The equations of motion for the ladder spin operator
$S_j^-\equiv S_j^x-iS_j^y$ and the longitudinal spin operator $S_j^z$
are obtained [3--6] from the Heisenberg equations for the spin operators,
with allowance made for the attenuation terms. This yields the equations
of motion
$$
\frac{dS_j^-}{dt} = - i
\left ( \om_0 + \xi_j^0 - i\Gm_2\right  ) S_j^- + f_j S_j^z +
i\; \frac{\om_D}{S} \; S_j^z S_j^- \; ,
$$
\be
\label{11}
\frac{dS_j^z}{dt} = - \;
\frac{1}{2} \; \left ( f_j^+ S_j^- + S_j^+ f_j \right ) -
\gm_1 \left ( S_j^z - \zeta \right ) \; ,
\ee
in which $\zeta$ is an equilibrium spin polarization and
$$
f_j = -\; \frac{i}{\hbar} \; \mu_0 H + \xi_j \; , \qquad
\Gm_2 = \gm_2 \left ( 1 - s^2 \right ) \; ,
$$
$$
\xi_j^0 =  \frac{1}{\hbar} \; \sum_{j(\neq i)}
\left ( a_{ij} S_j^z + c_{ij}^* S_j^- + c_{ij} S_j^+ \right ) \; ,
$$
$$
\xi_j =  \frac{i}{\hbar} \; \sum_{j(\neq i)}
\left ( 2 c_{ij} S_j^z -\; \frac{1}{2} \; a_{ij} S_j^- +
2b_{ij} S_j^+ \right ) \; .
$$
Here
\be
\label{12}
s  \equiv \frac{1}{SN} \; \sum_{j=1}^N < S_j^z >
\ee
is the reduced longitudinal spin polarization.

To characterize the level of coherence in spin motion, we consider the
total magnetodipole radiation intensity
\be
\label{13}
I(t) = I_{inc}(t) + I_{coh}(t) \; ,
\ee
which is separated into the incoherent and coherent parts,
\be
\label{14}
I_{inc} = \frac{2\mu_0^2}{3c^3} \;
\sum_j | < \ddot{\bS}_j > |^2 \; , \qquad
I_{coh} = \frac{2\mu_0^2}{3c^3} \; \sum_{i\neq j}
< \ddot{\bS}_i \ddot{\bS}_j>  \; ,
\ee
where the overdots imply the time differentiation. The level of dynamic spin
coherence is described by the {\it coherence factor}
\be
\label{15}
C(t) \equiv \frac{I_{coh}(t)}{I(t)} \; .
\ee

The system is, first, prepared in a nonequilibrium state, with a positive
initial spin polarization $s_0=s(0)$ in a positive external magnetic field
$B_0$. According to Eq. (2). molecular spins tend to reverse to negative
values corresponding to an equilibrium state.

\section{Collective dynamics}

We analyzed the system of equations (11) in two ways. One way is by
treating the spin variables $S_j^\al$ as operators, averaging Eqs. (11),
and employing the scale-separation approach [8,9]. The second way is by
considering $S_j^\al$ as classical variables and accomplishing direct computer
simulations of Eqs. (11). The scale-separation approach is more appropriate
for the initial stage of spin relaxation, when spins fluctuate chaotically and
their quantum nature is essential. At the stage of the coherent spin motion,
both ways give close results, though the direct computer simulation is more
accurate, exhibiting some fine-structure oscillations that are smoothed out in
the averaging technique.

Accomplishing numerical calculations for different system parameters, we set
the filling factor $\eta=1$ and take into account that for typical magnetic
nanomolecules $\gm_1\ll\gm_2$. By varying other parameters, we find that the
fastest spin reversal occurs for the largest $\om_0$ and $s_0$, but for the
smallest $\gm$ and $\om_D$. Dipole interactions stronger destroy spin coherence
for larger spins. For lower spins, the spin reversal is more pronounced. This
is illustrated in Fig. 1 for $S=10$ and $S=1/2$. In this and all following
figures, we present the results of the direct computer simulation with $N=125$
magnetic nanomolecules. The initial spin polarization (12) is taken as $s_0=0.9$
Time is measured in units of $T_2$ and all frequencies and damping parameters
in units of $\gm_2$. The resonance condition $\om_0=\om$ is assumed. The
anisotropy frequency is taken as $\om_D=20$.

Figures 2 to 4 show the temporal behavior of the coherence factor (15) for
the varying Zeeman frequency $\om_0$ and the resonator damping $\gm$. These
figures demonstrate that the collective spin motion is highly coherent. Such
a high level of coherence is due to the presence of the resonator, without
which spin dynamics could not be coherent [3--9], so that no superradiance
could exist [10].

In conclusion, we have considered spin dynamics in a strongly nonequilibrium
system of magnetic nanomolecules. By varying the system parameters, we find
the optimal configurations for the fastest spin reversal. For typical magnetic
nanomolecules, the reversed time can be of order $10^{-12}-10^{-11}$ s. Without
a resonator, the spin motion cannot be made coherent, provided no strong
transverse fields are imposed. But in the presence of the resonator, producing
a feedback field, spin dynamics can become essentially coherent. The ultrafast
self-organized spin reversal can be employed for information processing.

\vskip 5mm

{\bf Acknowledgement}

\vskip 3mm

The authors acknowledge financial support from the Russian
Foundation for Basic Research:  Grant 07-02-96026 (V.K.H. and
P.V.K.) and Grant 08-02-00118 (V.I.Y. and E.P.Y.).

\newpage

\begin{figure}[ht]
\begin{minipage}{7cm}
\includegraphics[width=7cm]{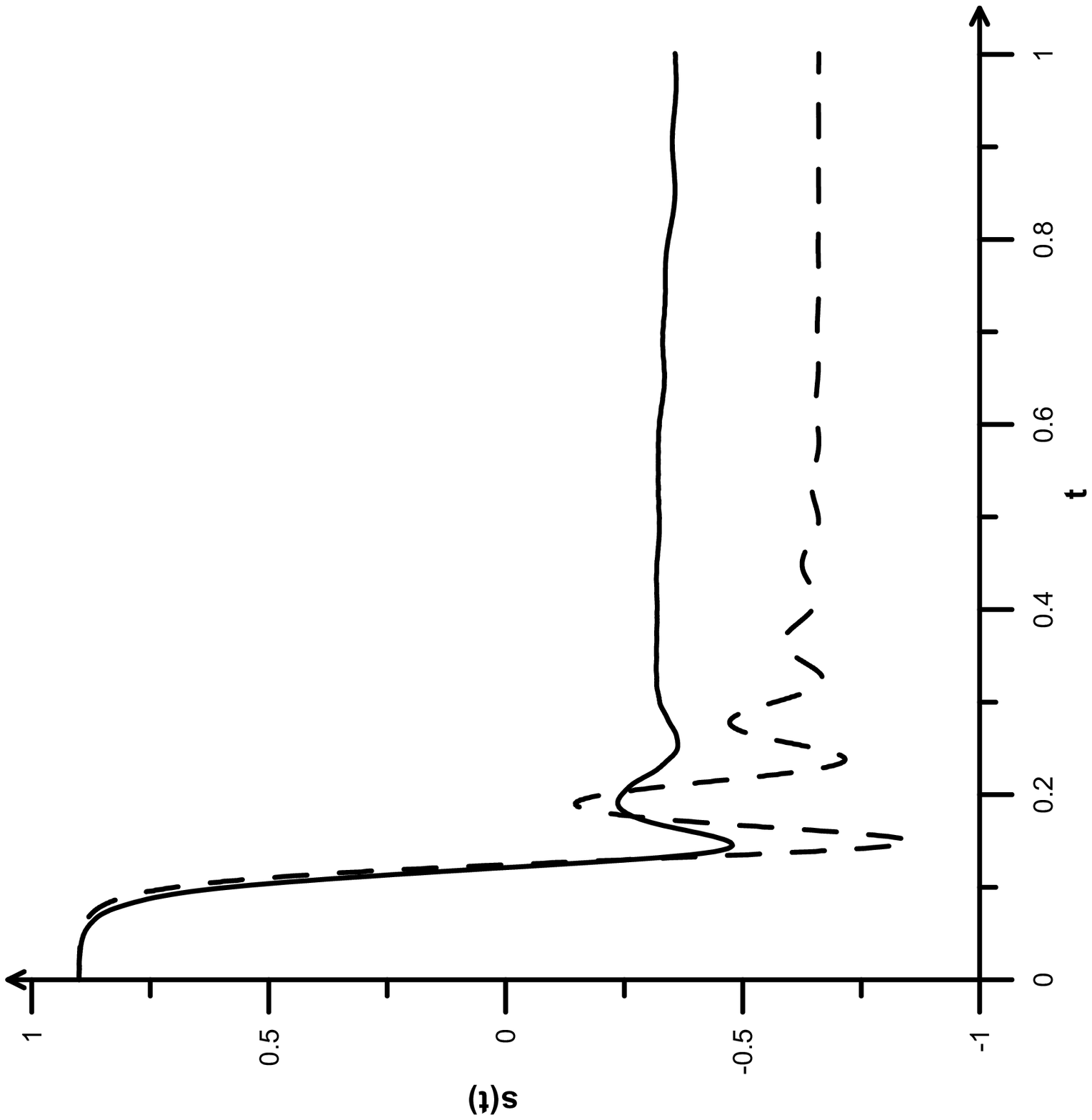}
\caption{Reduced spin polarization $s=s(t)$ as a function of
dimensionless time for $\om_0=2000$ and $\gm=10$, with the molecular spin
$S=10$ (solid line) and $S=1/2$ (dashed line).}
\end{minipage}\hspace{2cm}%
\begin{minipage}{7cm}
\includegraphics[width=7cm]{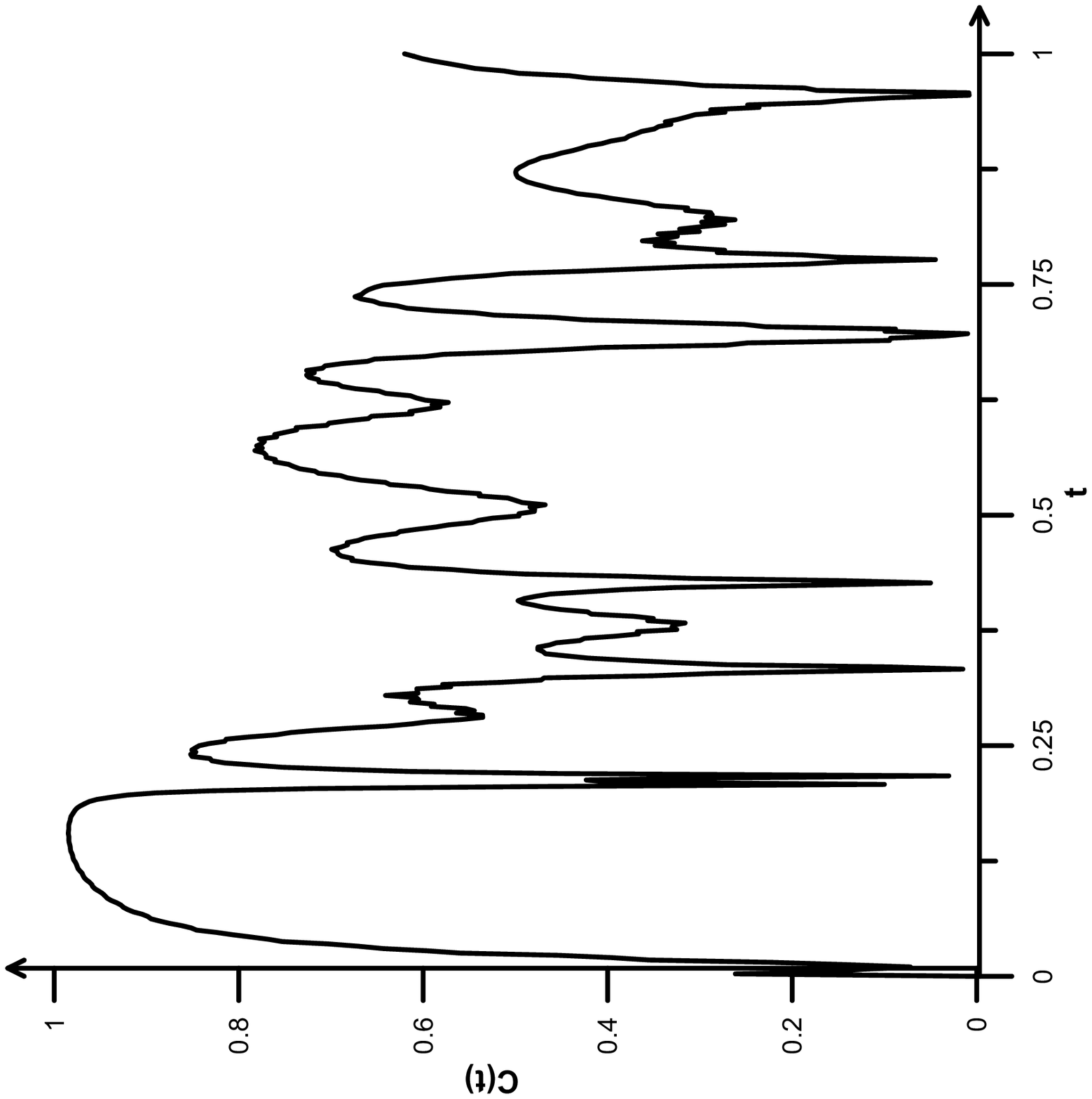}
\caption{Coherence factor (15) as a function of dimensionless time
for  $\om_0=1000$, $\gm=10$, and $S=10$.}
\end{minipage}
\end{figure}

\begin{figure}[ht]
\begin{minipage}{7cm}
\includegraphics[width=7cm]{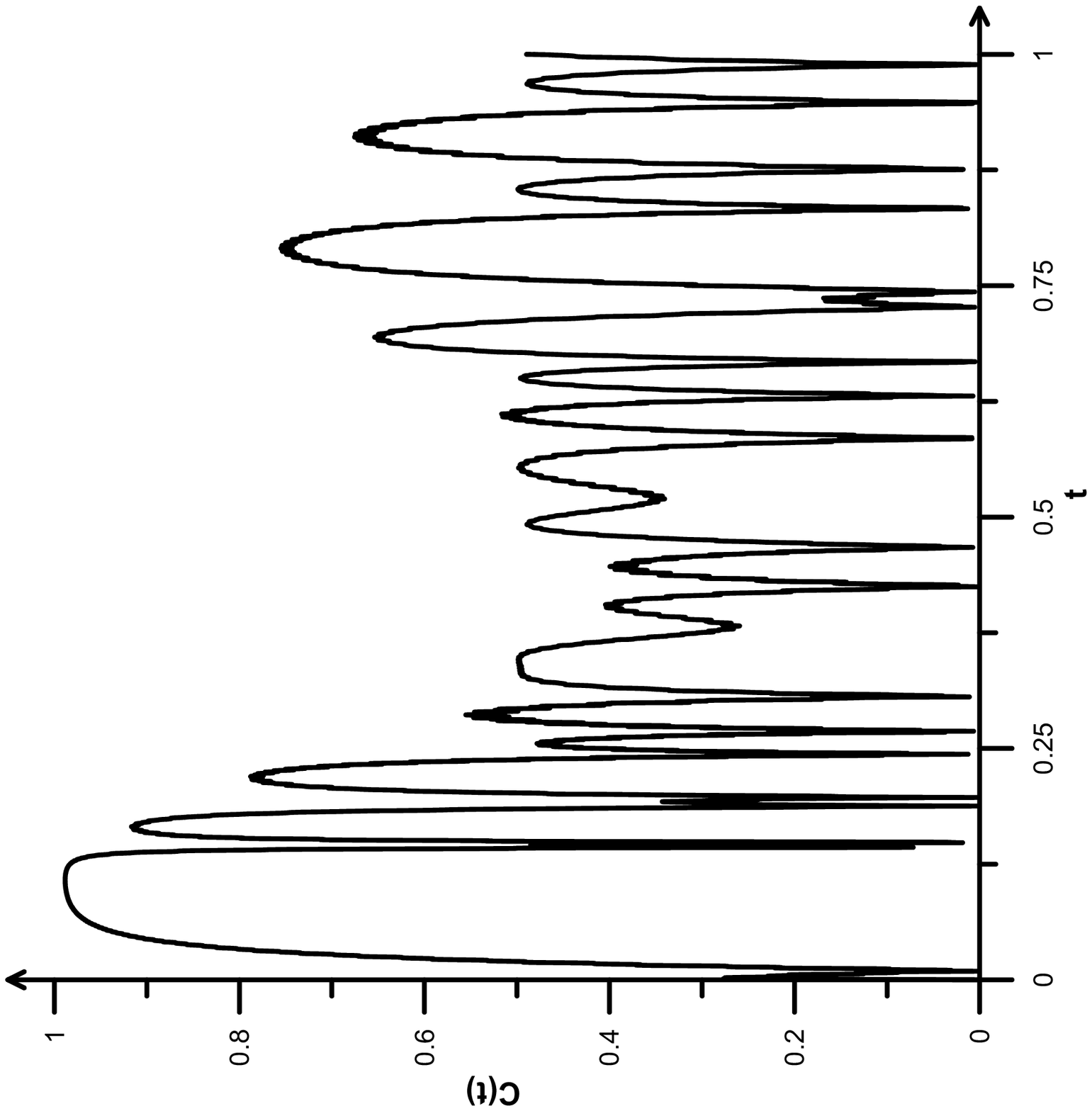}
\caption{Coherence factor (15) as a function of dimensionless time
for  $S=10$, $\gm=10$, but  $\om_0=2000$.}
\end{minipage}\hspace{2cm}%
\begin{minipage}{7cm}
\includegraphics[width=7cm]{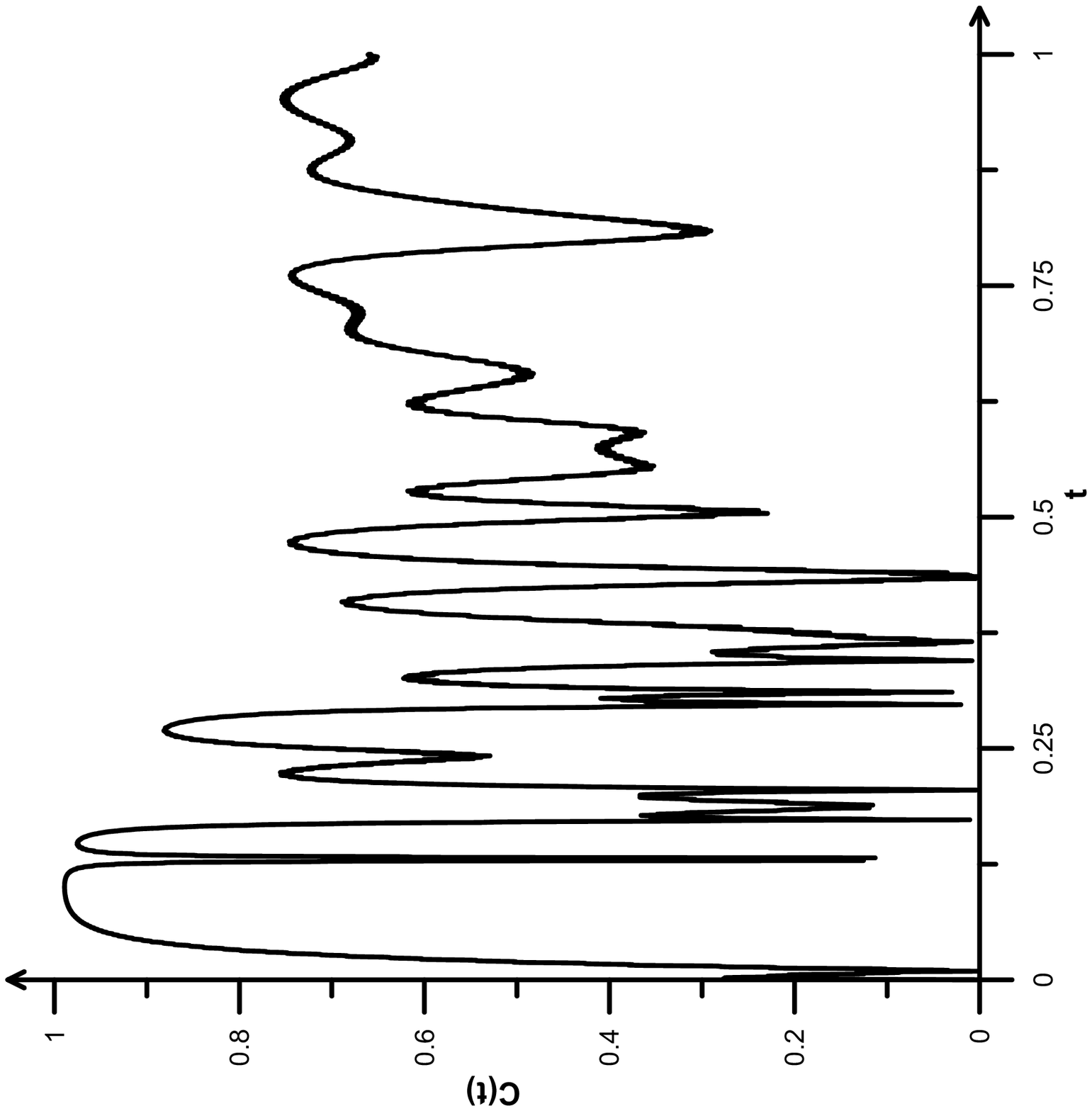}
\caption{Coherence factor (15) as a function of dimensionless time
for $S=10$, $\om_0=2000$, but $\gm=1$.}
\end{minipage}
\end{figure}

\end{document}